# Experimental study of the coherent A$_1$ phonons in Te with tailored femtosecond pulses

O V Misochko[1], M V Lebedev[1], H Schäfer[2], and T Dekorsy[2]

[1]Institute of Solid State Physics, Russian Academy of Sciences

142432 Chernogolovka, Moscow region, Russia

Email: misochko@issp.ac.ru

[2]Physics Department, Konstanz University, 78457 Konstanz, Germany

**Abstract**

We tailor the shape and phase of the pump pulse spectrum in order to study the coherent lattice dynamics in tellurium. Employing the coherent control via splitting the pump pulse into a two-pulse sequence, we show that the oscillations due to A$_1$ coherent phonons can be cancelled but not enhanced as compared to single pulse excitation. We further demonstrate that a decisive factor for the coherent phonon generation is the bandwidth of the pulse spectrum and not the steepness of the pulse envelope. We also observe that the coherent amplitude for long pump pulses decreases exponentially independent of the shape of the pulse spectrum. Finally, by varying the pulse chirp, we show that the coherent amplitude is independent of while the oscillation lifetime is dependent on the chirp sign.

Experimental study of the coherent $A_1$ phonons in Te with tailored femtosecond pulses

Soon after the first observation of coherent phonons (*i.e.,* lattice vibrations where all atoms move in phase with each other) in semimetals and narrow band semiconductors [1], Zeiger *et.al.* suggested a cogent explanation of the creation of lattice coherence in opaque materials called displacive excitation of coherent phonons (DECP) [2]. In this mechanism, after an ultrafast excitation, a significant fraction of the valence electrons are excited to higher electronic states. Such step-like excitation greatly reduces the attractive part of the interatomic potential and allows the atoms move with their thermal velocities towards a new equilibrium state. Some time later, it was suggested [3,4] that DECP is a particular (resonance) case of impulsive stimulated Raman scattering (ISRS), a mechanism which had been suggested earlier to explain the coherent vibrations of molecules and atoms in transparent crystals [5,6]. ISRS relies exclusively on the spectral bandwidth of ultrashort pulse [5-9] that exceeds the phonon frequency thus allowing the pulse to exert a temporally impulsive driving force on the atoms. Such force gives a kick to the atoms changing their thermal velocities. Since the lattice interacts with light only through electrons, the both mechanisms rely on the specific (deformation potential) electron-phonon coupling leaving electron-photon interaction to a great extent unspecified. Taking a simple analogy with a classical pendulum, one can relate non-resonant ISRS to kicking the pendulum (*i.e.,* applying a real force) and thereby to changing its kinetic energy, whereas DECP to modifying the potential energy of the pendulum (*i.e.,* applying an imaginary force). The main difference between the two models is the role of the electromagnetic field in the creation of the lattice coherence [7,9]. In DECP, any macroscopic variable affected by the optic field and coupled to the lattice can be used as the driving force provided that the variable can be switched in a time shorter than the phonon period. In this case, the light penetration depth defines both the phonon amplitude and wave vector, while the energy is supplied to the lattice through an (dominantly linear in intensity) absorption process. In resonant ISRS, where the energy is supplied to the lattice by a Stokes process, the electromagnetic field defines not only the phonon amplitude and wave vector (through the kinematic selection rules), but also the phonon

Experimental study of the coherent $A_1$ phonons in Te with tailored femtosecond pulses symmetry (through the symmetry selection rules). Thus, the differences between resonant ISRS and DECP include the possibility for the former of coherently exciting non-symmetric phonons and the requirement of phase matching based on kinematic constrains during the generation process [7].

In this paper, the goal is to study the role of the electromagnetic field in the coherent phonon generation. Since ultrashort laser pulses intrinsically consist of a broad range of frequency components, this provides the possibility of modifying the pulse spectrum through a controlled change of the amplitude and phase for each frequency. Such femtosecond pulse shaping has opened up a wide range of intriguing possibilities for controlling and manipulating lattice dynamics. The objective of studies in optical control over coherent phonons is to prepare a crystal in a nonequilibrium state in order to influence, and therefore to study, its evolution on femtosecond time scales. Pioneering experiments on the lattice coherent control were done in organic [10] and transparent crystals [11], and similar, though much simpler, studies have been conducted for opaque crystals at low [12,13] and high [14-17] excitation strength. We will use the pulse shaping and coherent control techniques to clarify the role of photons in the generation of lattice coherence. A favourable candidate to investigate the photon contribution in the creation of lattice coherence should be a simple crystal, which is relatively easy to handle, both theoretically and experimentally. For that reason, in our study single crystal of Te is used as a benchmark system as it has been extensively studied in the time-domain [1,2,18-24]. Its crystal structure $D_3^4$ consists of three-atom-per turn helices whose axes are arranged on a hexagonal lattice. Among six optical phonons, the fully symmetric $A_1$ phonon ($\Omega \approx 3.6$ THz) is a lattice mode for which the helical radius changes leaving the interhelical distance and *c*-axis spacing unchanged. Although all the optical phonons in Te have been excited coherently [18,20], the $A_1$ phonon mode is usually dominant as a result of large deformation potential for the symmetry preserving movements [24]. To further eliminate any contribution of unwanted optical phonons, we excite the coherent phonons in our study with a pump polarized along the trigonal axis thus

Experimental study of the coherent $A_1$ phonons in Te with tailored femtosecond pulses avoiding the generation of coherent phonons of lower symmetry [20,23]. Note that such geometry allows exciting coherent phonons through a change in the extraordinary part of the susceptibility $\chi_{ext}$, while probing the optical response through a change in the ordinary part of the susceptibility $\chi_{ord}$. Before describing our experimental results, we would like to stress that the task of settling the question as to which of the excitation mechanisms is the dominant one (or what is their interrelation) cannot be answered ultimately on the base of our experiments. The theoretical calculations in [2,19] support the DECP picture with respect to the fully symmetric $A_1$ mode, but the Raman framework or other mechanism based for example on the Dember field effect [21] are definitely required to account for the excitation of non-symmetry preserving coherent phonons.

Our laser system is based on a conventional Ti:sapphire laser which supplies 50 fs-800 nm pulses with the repetition rate of 82 MHz. Part of the beam was used as the pump pulse. The remaining part was used as the probe pulse, which is always unchirped, transform-limited pulse. The pump pulse was tailored with a pulse shaper comprising a spatial light modulator utilizing liquid crystals [25]. This approach to ultrafast pulse shaping involves a spatial Fourier transformation of the incident pulse to disperse the frequencies in space and modify the chosen frequency components selectively. A final recombination of all the frequencies into a single collimated beam results in the desired pulse shape. In our setup, a 1200 groves/mm grating spreads the pulse, so that each different spectral component maps onto a different spatial position. The collimating cylindrical lenses and grating pair are set up in a modified 4f strecher configuration (*f* being the focal length of the lenses) with 3 lenses instead of 1 for the two reasons: First, we are able to shorten the distance between the gratings to nearly 2f thus reducing the acoustic noise influences. Second, we obtain a better focus in the plane of the liquid crystals by taking the beam divergence into account. The liquid crystal double mask consisting of 320 stripes is placed that modulate the spectrum. Both the tailored pump and the transform-limited probe were focused by a single 10 cm lens onto the sample. At the sample

Experimental study of the coherent $A_1$ phonons in Te with tailored femtosecond pulses surface they were linearly (orthogonal to each other) polarized in the plane of incidence. The transient reflectivity was measured with a fast scanning delay line. A polarizer was inserted between the sample and the *p-i-n* detector to prevent scattered pump light from reaching the detector and to ensure that only properly polarized light is measured. To establish time zero, the roles of the pump and probe were reversed and the zero delay time was precisely determined by locating the point of mirror symmetry. The error in establishing time zero was less than 5 fs. The pulse duration of the probe was measured with a standard autocorrelator to get transform-limited pulses at the output. A cross correlation between pump an probe was used to calibrate the pulse shaper phase modulation, concerning the chirp of additional glass in the pump line. The cross-correlation of the tailored pump pulse with the bandwidth limited probe pulse was taken each time to confirm the correct modulation.

Given that tellurium at room temperature is a narrow band semiconductor, after the ultrafast excitation at λ=800 nm (1.55 eV), both coherent phonons and free carriers are excited simultaneously [1,2]. As a result, the fast oscillations in transient reflectivity are observed superimposed onto the long-time electronic decay. To separate the oscillatory $\left(\frac{\Delta R}{R_0}\right)_{osc}$ and electronic $\left(\frac{\Delta R}{R_0}\right)_{el}$ contributions, we will usually fit the signal $\frac{\Delta R}{R_0}$ to

$$H(t)\left[\left(\frac{\Delta R}{R_0}\right)_{el}\exp(-t/\tau_{el})+\left(\frac{\Delta R}{R_0}\right)_{osc}\exp(-t/\tau_{osc})\sin(\Omega t+\varphi)\right],$$

where $H(t)$ is the Heaviside function convoluted with the pump pulse, $\Omega$ and $\tau_{osc}$ are the frequency and lifetime of oscillations, while $\tau_{el}$ is the electronic relaxation time.

We first split the 50 fs transform-limited pulse into two identical Gaussian pulses with a different interpulse separation γ. Figure 1(a), where the four traces are designed to illustrate the changes in the response when the interpulse separation γ is varied, shows the transient reflectivity of Te excited with such two-pulse sequence (each pulse is unchirped). The traces

Experimental study of the coherent $A_1$ phonons in Te with tailored femtosecond pulses from the bottom to the top correspond to the interpulse separation γ of 80, 140, 200, and 280 fs, respectively. If the total electronic contribution is always the sum of the signals resulting from each pulse independently, the oscillatory component exhibits interferences resulting in a variation of the total amplitude as a function of γ. As shown in figure 1(b) depending on the interpulse separation, the oscillatory component can be almost completely cancelled but never enhanced as compared to single pulse excitation (that is $\left(\frac{\Delta R}{R_0}\right)_{osc}^{\gamma=0} \geq \left(\frac{\Delta R}{R_0}\right)_{osc}^{\gamma \neq 0}$ ). On the other hand, at the time delay γ corresponding to the arrival of the second pump pulse, a time delay of γ=nT between the two pump pulses, where n is an integer and T is the phonon period, results in enhancement while a time delay of (n+1/2)T results in cancellation of the oscillations created by the first pulse, *i.e.,* achieving oscillation cancellation or enhancement is sensitive to the arrival time of the second pump pulse relative to the first. That is the coherent oscillations created by the *first* pulse can be either cancelled or enhanced, the latter option realized with a restriction – the resulting amplitude cannot be larger than that created by the single pulse of the same energy as two pulse sequence. Thus, splitting the excitation energy (or the number of photons) between multiple pulses provide selectivity, but not increase in the resulting coherent amplitude suggesting that the lattice excitation is proportional to the total energy (or intensity) of the laser pulse.

In contrast to high excitation case [14-16], our experiments show that the oscillations under low excitation can be cancelled *exactly* at the maximum displacement when the coherent amplitude has a different sign compared to that at time zero. Indeed, the 140 fs pulse separation that achieves almost full cancellation as shown in figure 1(a) coincides with the time to reach the maximum displacement (inner classical turning point) in the single pulse pump case, since the oscillation period is approximately 280 fs and the oscillation starts at the outer turning point [26]. Similar cancellation of the coherent vibrations at the maximum displacement has been observed in bismuth [13] and GaAs [12] for weak excitation strength.

Experimental study of the coherent $A_1$ phonons in Te with tailored femtosecond pulses

In the low excitation regime, interaction between an ultrashort pulse and a crystal can be explained adequately by perturbation theory. For one-photon transitions, this means that the interaction varies linearly with the exciting field. Thus, the crystal response to a two-pulse sequence can be described either by summing the contributions of each pulse, or by considering directly the effect of the total exciting field. Although not correct in general, in our case the coherent control can be well understood as the sum of two sets of coherent phonons whose motion is initiated at different times and that then interfere (this description ignore the interaction of the second pulse with already excited coherent movement). This model was first introduced in [13] to describe the coherent control of $A_{1g}$ phonons in Bi at low excitation strength. Nevertheless, what we observe in fact is not the simple sum of two single-pulse signals $R_1^{t=0}(t) + R_2^{t=\gamma}(t)$ with $\gamma$ being control parameter, but the transient reflectivity change $\chi_{tot}(t)$ that derives from the sum of two oscillating susceptibilities $\chi_1^{t=0}(t) + \chi_2^{t=\gamma}(t)$. Only because our square root detectors measure essentially, $|\chi_{tot}(t)|^2$, in which cross-term with oscillatory component vanishes, we cannot see the difference between $R_1^{t=0}(t) + R_2^{t=\gamma}(t)$ and $\chi_1^{t=0}(t) + \chi_2^{t=\gamma}(t)$.

Given that in Te it is the nonequilibrium carrier distribution that drives coherent phonons, the process of coherent control can be better thought of in the following way [15]: The first pump pulse creates a new potential in which the atoms will move. Initially displaced from the newly established equilibrium configuration, the lattice reaches this configuration in approximately one quarter of a phonon period, but due to inertia the atoms have momentum at that point and continue to move. When the atoms reach the classical turning point (located on the opposite slope of the potential) of their motion, a second pump pulse can excite the precise density of carriers to shift the equilibrium position to the current position of the atoms, stopping the oscillatory motion. Because the photoexcitation of additional carriers can only increase (or decrease) the equilibrium helical radius [26], the vibration can only be stopped at the maximum displacement (when atoms are located on the opposite slope of the potential as compared to that from where the movement starts) as it is observed in our experiments.

Second set of experiments reported in this work will focus on the question what is more important for coherent phonon generation: pulse spectrum or steepness of the leading edge of the pulse envelope? To separate these two effects, we use the pulses with different almost rectangular spectra that form a kind of sinc function in the time domain. Figure 2 shows the oscillations excited by such tailored pulses. The coherent oscillations for two different spectra of the pump pulse were studied: the rectangular spectrum with bandwidth smaller that phonon frequency, and the spectrum with a hole in the middle (in this case the overall bandwidth is larger that phonon frequency). As expected, relatively large oscillatory amplitude is observed for the case with the pulse spectrum containing phonon frequency. In contrast, the oscillations almost disappear (their amplitudes are 6-7 times smaller) when the spectrum is narrower than the phonon frequency. Note that the electronic contributions $\left(\frac{\Delta R}{R_0}\right)_{el}^{max}$ in the both cases almost coincide and the different oscillatory amplitudes unlikely can be *quantitatively* explained by the constructive interference resulting from two-pulse sequence in the case of spectrum with a hole (even if the nT criterion for enhancement is satisfied it cannot result in sevenfold enhancement). The vanishing oscillatory amplitude $\left(\frac{\Delta R}{R_0}\right)_{osc}$ in the case of single sinc pulse challenges to some extent the applicability of DECP as the primary excitation mechanism for fully symmetric phonons. Indeed, in DECP picture, a crucial requirement for coherent phonon generation is that the changes induced by the exciting pulse are abrupt, that is the driving force is step-like. The abruptness can be related to a steepness of the leading edge of the pulse envelope. Therefore, approximating the sinc pulse by a triangular function in the time-domain

$$I(t) = \Delta\left(\frac{t}{\tau}\right) = \begin{cases} 1 - \frac{|t|}{\tau} & \text{if } \tau < t \leq \tau \\ 0 & \text{otherwise} \end{cases}$$

Experimental study of the coherent $A_1$ phonons in Te with tailored femtosecond pulses one can see that the pulse with a broader spectrum has a steeper leading edge than each of the two triangular pulses formed from the spectrum with a hole in the middle because the bandwidth of each pulse is narrower than that of the single (broader spectrum) pulse [27]. Thus, this observation seems to suggest that a crucial condition to effectively create the lattice coherence is the bandwidth of pulse spectrum. Such a condition implies that the pump process can be somehow related to biharmonic pumping (either two photon absorption, or scattering) that involves at least two field-matter interactions necessary for the energy transfer and the coherence creation. The important point to note is that in this case the phases of the lattice eigenstates are set by the phase of the relevant spectral components of the electromagnetic field, and thus appropriate manipulation of the optical phase can produce radically different lattice states.

In order to further clarify the role of pulse envelope, we compare coherent phonons excited with sawtooth pulses (*i.e.,* pulses having a linear rise on the leading edge and a virtually instantaneous fall on the trailing edge, or conversely, a virtually instantaneous rise and a linear fall). These two types of the sawtooth pulse shown in figure 3 together with a 50 fs transform limited and a 1 ps positively chirped pulse, are ideally suited to study the effects of leading edge since their temporal profiles are inverted in time. In addition, their spectra (almost identical) were chosen to be significantly narrow than phonon frequency, see figure 4. Unexpectedly, coherent oscillations $\left(\frac{\Delta R}{R_0}\right)_{osc}$ excited with such sawtooth pulses are almost equal, with relatively large oscillatory amplitude reaching one half of that observed for the 50 fs transform limited pump pulse of the same energy, see figure 5. The comparable oscillatory amplitudes indicate that coherent phonon generation seems to be independent of the steepness of leading edge. However, the amplitude magnitudes themselves question the importance of pulse bandwidth. Moreover, the result for the sawtooth pulses is, at first glance, in conflict with that

Experimental study of the coherent $A_1$ phonons in Te with tailored femtosecond pulses obtained for the sinc pulses since the spectral bandwidth of the used sawtooth pulses is much smaller than the phonon frequency. One way of resolving the discrepancy between the sinc and sawtooth pulses is to assume that the coherent oscillations observed with sawtooth pulses are due the spectral sidebands intrinsic for the pulses of such temporal profile. Indeed, a closer inspection of the temporal profile reveals that a linear rise (drop) in the both cases is slightly modulated with the ≈80 fs period, multiples of which are comparable with the inverse of phonon frequency. In this case the number and the temporal separation of the subpeaks, which coincide for both types of the used sawtooth pulses, can essentially define the resulting amplitude.

Having done with coherent control experiments, we next focus our attention on the question how the duration of a single laser pulse affects the creation of lattice coherence? Even though it is well known that the pulse duration has to be shorter than the phonon period (or equivalently the spectral width of the pump pulse has to be broader than the phonon frequency) in order to achieve efficient coherent excitation, the experimental demonstration of this requirement has been not realized yet, at least for opaque materials. In order to study the pulse duration effects, we make the Gaussian and rectangular pulses of different bandwidths and trace the parameters of transient response as a function of the pulse duration which is inversely proportional to the bandwidth. Figure 6 reflects the influence of the pulse stretching on the efficiency of lattice excitation for the two different spectral line shapes. For a Gaussian pulse, coherent amplitude $\left(\frac{\Delta R}{R_0}\right)_{osc}$ decreases exponentially with pulse duration, see figure 6(a). The characteristic pulse duration for the amplitude reduction is 75fs ≈ T/4 as obtained from the fit to an exponential function. There is a simple physical reason behind this relation, which can be explained by a model of driven pendulum. Assume the pendulum has a period of T=1/Ω and initially is in its equilibrium position. An external perturbation, in our case linked to the electromagnetic field, can accelerate it most effectively in the first quarter of its period, *i.e.,*

Experimental study of the coherent $A_1$ phonons in Te with tailored femtosecond pulses

$\tau=T/4$. If the perturbation is shorter or longer than T/4, the pendulum will not achieve the optimal energy and its amplitude will decrease.

For a Gaussian pulse, we identify no detectable phase lag of the oscillations most likely due to the fact that damping is significantly smaller than any of the pulse durations. In contrast, for a rectangular pulse

$$I(t) = \text{rect}\left(\frac{t}{\tau}\right) = \begin{cases} 1 & \text{if } -\frac{1}{2}\tau < t \leq \frac{1}{2}\tau \\ 0 & \text{otherwise} \end{cases}$$

the modulus of oscillatory amplitude $\left|\left(\frac{\Delta R}{R_0}\right)_{osc}\right|$ is not a smooth function of the pulse duration. It follows the regular sinc function ($=\frac{\sin \Omega \tau}{\Omega \tau}$) dependence changing the sign at $\tau = \frac{1}{\Omega}$ with $\Omega$ being the phonon frequency. When the pulse duration matches phonon period the amplitude changes its sign. This means that the phase experiences a $\pi$ jump at this duration as observed in our experiments and shown in the inset of figure 6(b).

It is interesting to note that the electronic contribution $\left(\frac{\Delta R}{R_0}\right)_{el}$ is independent of pulse duration as it could be also inferred from coherent control experiments. Since in our experiments the stretching of the pulses is achieved under the condition that the energy of the pulse is approximately conserved (as the mask covers a broader part of the spectrum of the initial transform limited pulse, the pulse becomes less stretched and its intensity increases) we conclude that coherent amplitude is proportional to pulse intensity, whereas non-equilibrium carrier concentration can be proportional to pulse energy. In other words, long and short pulses of equal energy excite the same density of electrons, leading to the same displaced equilibrium position, but long pulses excite smaller amplitude coherent phonons than short pulses.

In an attempt to separate the effects of energy and intensity of the pump pulse on the coherent phonon generation, we measure pump dependence for fixed pulse duration. To this end, we vary the average power of the train of the 50 fs transform limited pulses and trace the amplitude $\left(\frac{\Delta R}{R_0}\right)_{osc}$ and lifetime $\tau_{osc}$ (inversely proportional to dephasing rate) of the oscillations, as well as the electronic contribution $\left(\frac{\Delta R}{R_0}\right)_{el}$. As can be seen from figure 7, both the amplitude and dephasing rate increases approximately linear with pump power. The electronic contribution $\left(\frac{\Delta R}{R_0}\right)_{el}$ (not shown in the figure) is also linear and scales with the oscillatory amplitude $\left(\frac{\Delta R}{R_0}\right)_{osc}$. This linear dependence is consistent with the expectation that both the density of the nonequlibrium carrier and the oscillation amplitude scale with the mean number of photons in the pump pulse. It is interesting to note that a linear extrapolation to zero coherent amplitude suggests the existence of a threshold, which is expected for a nonlinear process like two photon absorption or stimulated Raman scattering. This threshold may indicate that linear absorption is not the only process by which the pump is absorbed in our crystal. To remove this threshold behaviour one has to assume that the pump power dependence is described by a nonlinear function as shown in the insert of figure 7(a).

The decrease in lifetime with increasing pump power, as shown in figure 7 (b), can be due to enhanced phonon-phonon and electron-phonon interaction resulting from additional hot carriers and hot (noncoherent) phonons that are produced by more intense pump pulses. We also observed a decrease in the oscillation frequency (not shown) for a higher pump power. This softening resulting from dynamical screening has been extensively discussed and explained in a number of papers [15,18-20]; therefore we will avoid discussion of this aspect of our results that simply verify previous observations.

Finally, we study the influence of linear chirp as we scan different bandwidths of the pump pulse (in our previous experiments the chirp was set to zero, *i.e.,* only the amplitude but not the

Experimental study of the coherent $A_1$ phonons in Te with tailored femtosecond pulses phase of electromagnetic field was modulated). In this last set of experiments, we vary not only the magnitude, but also the direction of chirp. In such linearly chirped pulses, the phase of each frequency component varies linearly in time. By using either positively (blue follows red) or negatively (red follows blue) chirped pulses, the arrival time of each frequency component at the crystal can be controlled. In order to study chirp dependence, we measure the transient reflectivity with such positively and negatively chirped pulses. For a larger absolute value of linear chirp and therefore for longer pulse duration, the electronic contribution $\left(\frac{\Delta R}{R_0}\right)_{el}$ and oscillation frequency $\Omega$ remain essentially unchanged. On the other hand, the larger is the chirp, the smaller is the oscillation amplitude. More exactly, the oscillatory contribution $\left(\frac{\Delta R}{R_0}\right)_{osc}$, shown in figure 8, continuously disappears for a larger chirp with a rate that is independent of the chirp sign. By comparing this decrease with the changes observed for different bandwidth, unchirped pulses, we conclude that this reduction is trivial being simply due to a larger duration of the chirped pulses. Note that the observed insensitivity to the chirp sign at low excitation strength is in striking contrast to the case of amplified pulses where the strong asymmetry for positively and negatively chirped pulses has been recently observed [23]. Nevertheless, lifetimes of the oscillations do show an asymmetry in the dependence on the chirp sign indicating that the oscillations created with positively chirped pulses live longer than those generated with negatively chirped ones. Given that the same durations for positively and negatively chirped pulses yield the asymmetry of lifetimes, such behaviour cannot be explained by the temporal broadening only. Relaying on a similar observation in molecules [28], we tentatively ascribe this asymmetry to different contributions of the excited and ground electronic states involved in the creation of lattice coherence. These different contributions may be due to the fact that by tailoring the spectral phase of the pulse, we control the interference between the various spectral components, leading to selective population of given phonon energy levels.

Experimental study of the coherent $A_1$ phonons in Te with tailored femtosecond pulses

In summary, our experiments are the first demonstration of a pulse-shaping strategy where a complex shape (in phase and amplitude) has been applied to study the coherent lattice dynamics in simple, opaque crystal. The experiments have demonstrated that in Te a necessary condition for the effective coherent $A_1$ phonon generation appears to be the bandwidth of the pump pulse. This fact combined with a threshold in the dependence of coherent amplitude on excitation strength suggests that both linear and nonlinear interactions between the pump and the crystal are important. By coherent control experiments, we have proved that coherent amplitude can be cancelled but not enhanced if the excitation energy is split into two pulses. The oscillations are stopped exactly at the maximum displacement corresponding to the inner turning point, or for time delays of n+1/2 phonon periods between successive pump pulses. Next, we have shown that an increase in the pulse duration results in an exponential decrease of coherent amplitude independent of the pulse envelope. However, it turns out that for the pulses with a rectangular envelope, the modulus of amplitude is not a smooth function and the phase of coherent phonons experiences a π-jump when the pulse duration matches phonon period. In addition, we observe that long and short pulses of equal energy excite the same density of electrons, leading to the same displaced equilibrium position, however long pulses excite smaller amplitude coherent phonons than short pulses. Finally, varying chirp and its sign, we have established that for low excitation strength the chirp sign does not affect the oscillation amplitude but brings on an asymmetry in the dependence of oscillation lifetime on the chirp sign. The results obtained reveal a wealth of information needed for a better understanding of coherent lattice dynamics of opaque crystals.

This work was supported in part by the Deutsche Forschungsgemeinschaft (DE567/9) and by the Russian Foundation for Basic Research (06-02-16186 and 07-02-00148).

Experimental study of the coherent $A_1$ phonons in Te with tailored femtosecond pulses

**REFERENCES**

1. Cheng T K, Vidal J, Zeiger H J, Dresselhaus G, Dresselhaus M S, and Ippen E P 1991 *App. Phys. Lett,* **59** 1923

2. Zeiger H J, Vidal J, Cheng T K, Ippen E P, Dresselhaus G, and Dresselhaus M S 1992 *Phys. Rev.* B **45** 768

3. Garrett G A, Albrecht T F, Whitaker J F, and Merlin R 1996 *Phys. Rev. Lett.* **77** 3661

4. Stevens T E, Kuhl J, and Merlin R 2002 *Phys. Rev.* B **65** 144304

5. Chesnoy J, and Mokhtari A 1988 *Phys.Rev.* A **38** 3566

6. For a review, see e.g Dhar L, Rogers J A, and Nelson K A 1994 *Chem. Rev.* **94** 167

7. Merlin R 1997 *Solid State Commun.* **102** 207

8. Dekorsy T, Cho G C, and Kurz H 2000 in *Light Scattering in Solids VIII*, eds.M. Cardona and G. Güntherodt, Springer, Berlin, 169

9. Misochko O V 2001 *Journal of Experimental and Theoretical Physics*, **92** 246 (translated from 2001 *ZhETF* **119** 285)

10. Weiner A M, Leaird D E, Wiederrecht G P, and Nelson K A 1990 *Science* **247** 1317

11. Wefers M M, Kawashima H, and Nelson K A 1996 *J. Phys. Chem. Sol.* **57** 1425

12. Dekorsy T, Kutt W, Pfeifer T, and Kurz H 1993 *Europhys. Lett.* **23** 223

13. Hase M, Mizoguchi K, Harima H, Nakashima S, Tani M, Sakai K, and Hangyo M 1996 *App. Phys. Lett.* **69** 2474

14. DeCamp M F, Reis D A, Bucksbaum P H and Merlin R 2001 *Phys. Rev.* B **64** 092301

15. Roeser C A D, Kandyla M, Mendioroz A and Mazur E 2004 *Phys. Rev.* B **70** 212302

16. Murray E D, Fritz D M, Wahlstrand J K, Fahy S, and. Reis D A 2005 *Phys.Rev*. B **72**, 060301 (R)

17. Misochko O V, Lu R, Hase M, and Kitajima M 2007 *Journal of Experimental and Theoretical Physics* **104** 245 (translated from 2007 *Zh.Exp.Teor.Fiz.* **131** 275)

18. Hunsche S, Wienecke K, Dekorsy T, and Kurz H 1995 *Phys.Rev.Lett.* **75** 1815

Experimental study of the coherent $A_1$ phonons in Te with tailored femtosecond pulses

19. Tangney P, and Fahy S 2002 *Phys. Rev.* B **65** 054302
20. Misochko O V, Lebedev M V, and Dekorsy T 2005 *J.Phys.:Condens. Matter* **17** 3015
21. Dekorsy T, Auer H, Bakker H J, Roskos H G, and Kurz H 1996 *Phys. Rev.* B **53** 4005
22. Hunsche S, Wienecke K, and Kurz H 1996 *Appl. Phys.* A **62** 502
23. Misochko O V, Dekorsy T, Andreev S V, Kompanets V O, Matveets Yu A, Stepanov A G, and Chekalin S V 2007 *App. Phys. Lett.* **90** 071901
24. Kudryashov S I, Kandyla M, Roeser C A D, and Mazur E 2007 *Phys. Rev.* B **75** 08520
25. Lee K G, Kim D S, Yee K J, and Lee H S 2006 *Phys. Rev.* B **74** 113201
26. In tellurium, density functional theory calculations [19] show that transferring electrons from the valence band to the conduction band changes the potential surface so that a larger equilibrium helical radius is established. Nevertheless, experimental studies disagree on this point, suggesting either expansion [10] or contraction [22] of the helical radius after the photoexcitation.
27. Note, however, that the spectral components of the two spectral parts separated by the hole might interfere and cause therefore a steep component in the time domain.
28. Bardeen C J, Wang Q, and Shank C V 1995 *Phys. Rev. Lett.* **75** 3410

Experimental study of the coherent A$_1$ phonons in Te with tailored femtosecond pulses

**FIGURE CAPTIONS**

Figure 1. Differential reflectivity in Te versus time delay for two-pulse excitation (a). The average incident power of the pump pulses is 50 mW. The interpulse separation γ is indicated for each trace displaced vertically for clarity. (b) Coherent amplitude as a function of the interpulse separation γ.

Figure 2. Differential reflectivity in Te versus time delay for the pump pulses with rectangular spectra. Their spectra (offset for clarity) are schematically shown in the inset. The dashed (red) trace and the solid (blue) trace correspond to the pump pulses whose bandwidths are correspondingly smaller or larger than phonon frequency.

Figure 3. Temporal profiles of the sawtooth pulses. For a comparison purpose, temporal profiles of the 50 fs transform limited pulse and of the 1 ps positively chirped pulse of the same energy are also shown.

Figure 4. Calculated spectra of the pulses shown in figure 3, the spectra of the different sawtooth pulses are overlapped.

Figure 5. Differential reflectivity versus time delay for the pulses shown in figure 3. The zero delay point is not calibrated.

Figure 6. Amplitude and phase of the coherent A$_1$ oscillations as a function of the pulse duration for the Gaussian (a) and rectangular (b) pulses. Dashed lines are fits using exponential and sinc functions, respectively. The insets show the oscillatory part at chosen time delays for the pulses with duration shorter (blue, solid line) and longer (red, dashed line) than phonon period. In the left inset the traces are offset for clarity.

Figure 7. Pump power dependence of the coherent amplitude (a) and of the oscillation lifetime (b). Dashed blue lines are linear fits, red dotted line is a non-linear fit.

Figure 8. Chirp dependence of the coherent amplitude (a), and of the oscillation lifetime (b). Open (blue) and closed (red) symbols are for negatively and positively chirped pulses, respectively. Dashed lines are linear fits.

Experimental study of the coherent $A_1$ phonons in Te with tailored femtosecond pulses

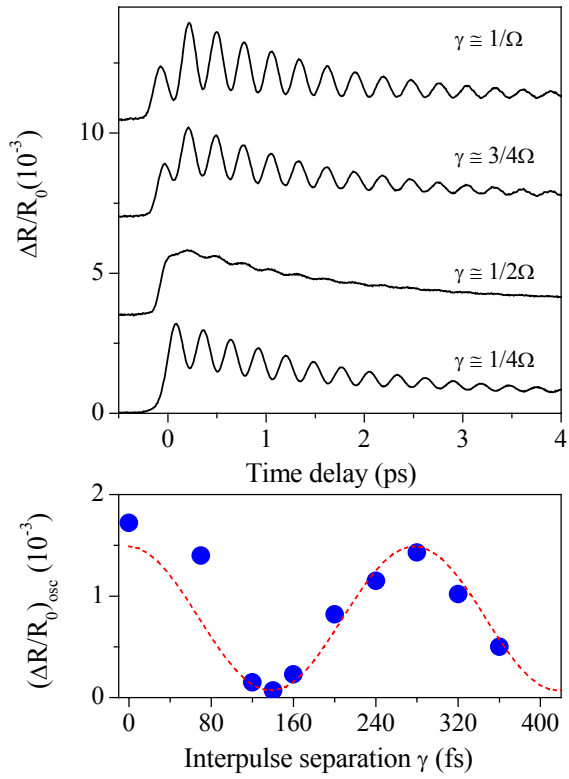

Figure 1.

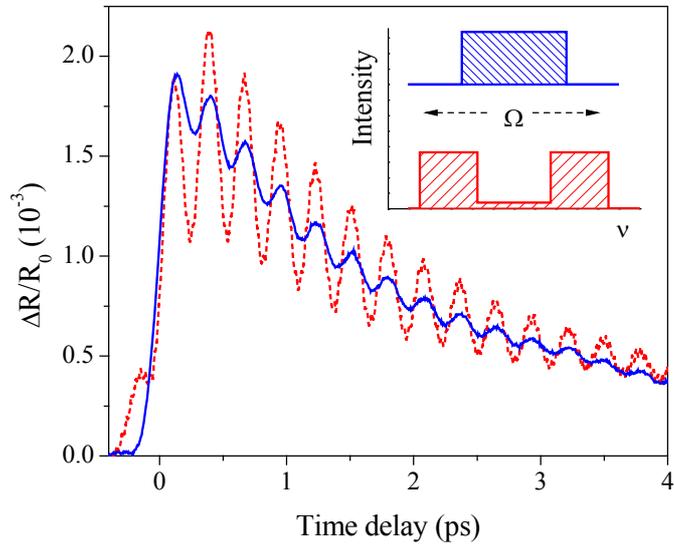

Figure 2.

Experimental study of the coherent $A_1$ phonons in Te with tailored femtosecond pulses

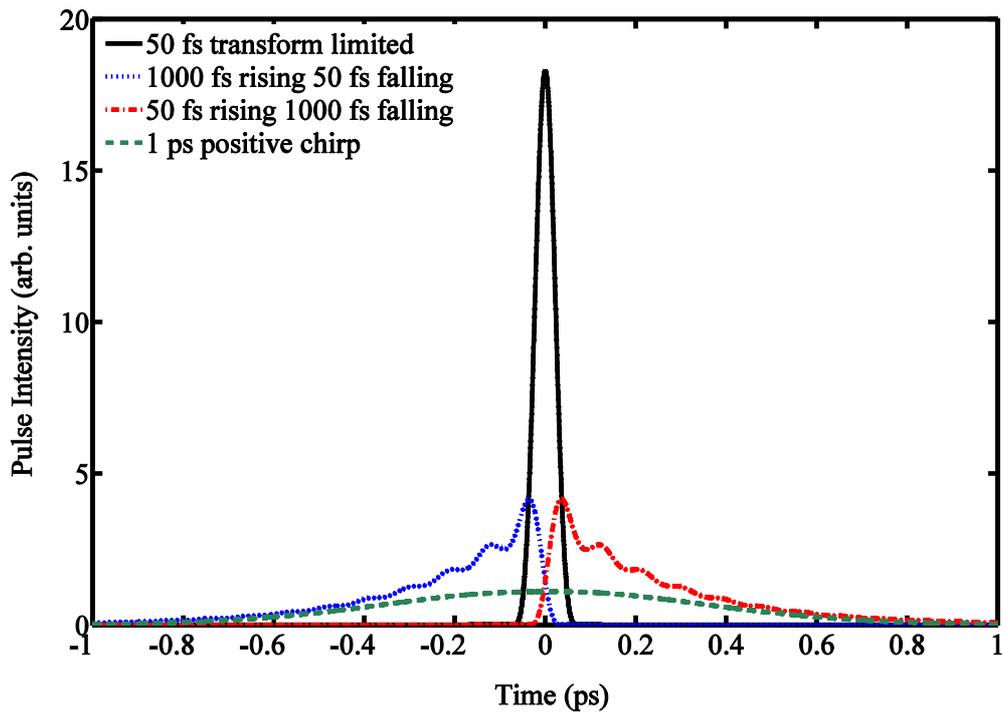

Figure 4.

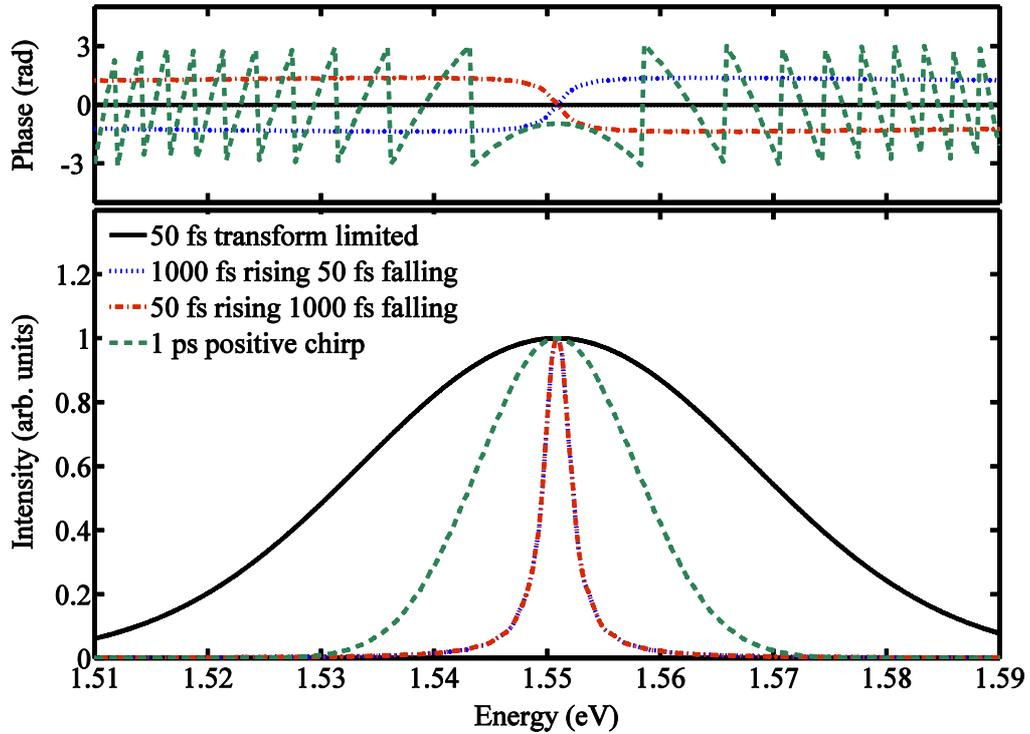

Figure 6.

Experimental study of the coherent $A_1$ phonons in Te with tailored femtosecond pulses

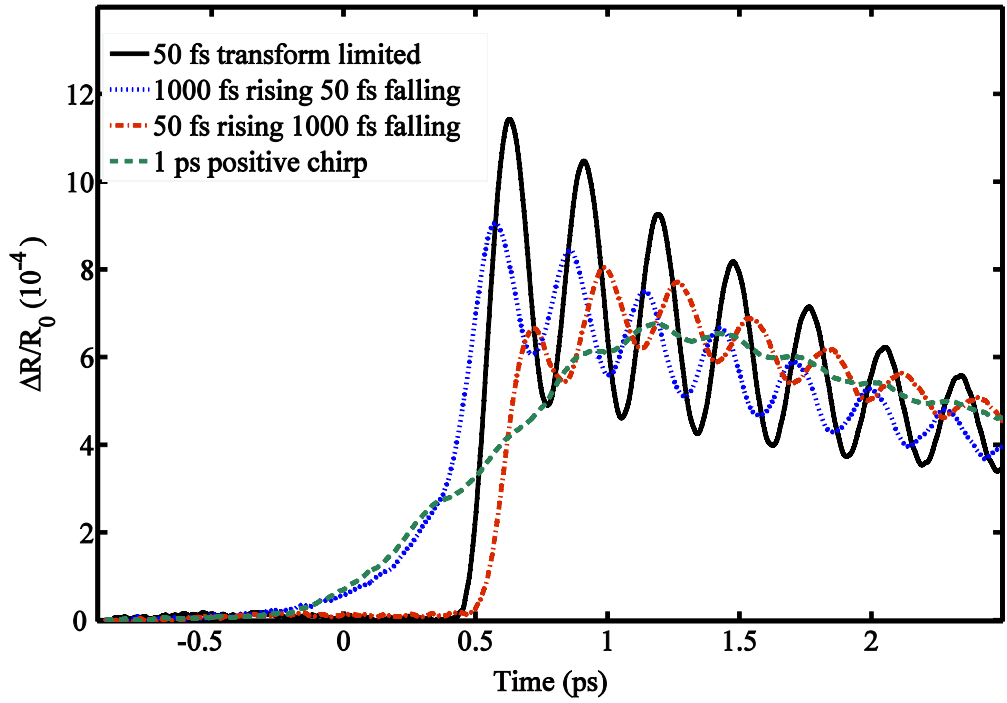

Figure 5.

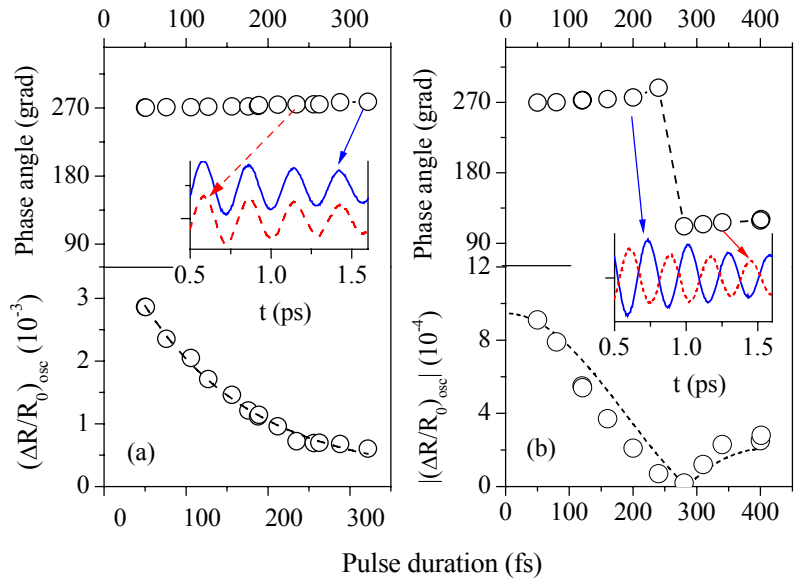

Figure 6.

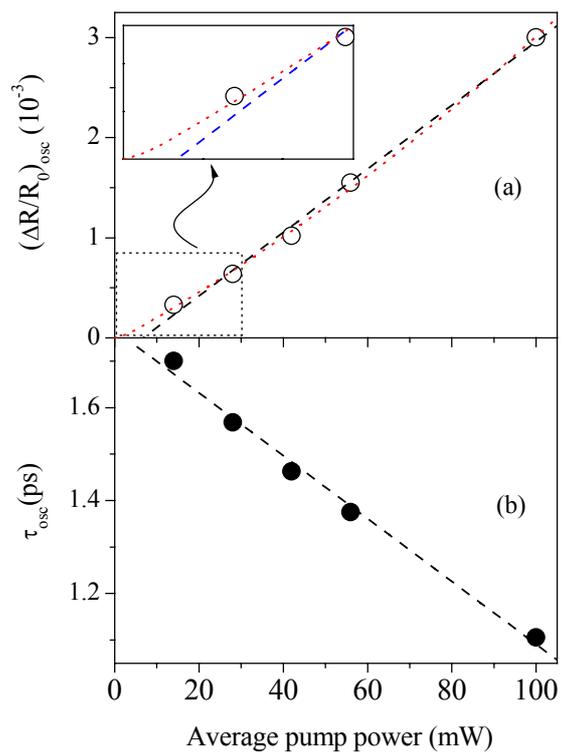

Figure 7.

Experimental study of the coherent $A_1$ phonons in Te with tailored femtosecond pulses

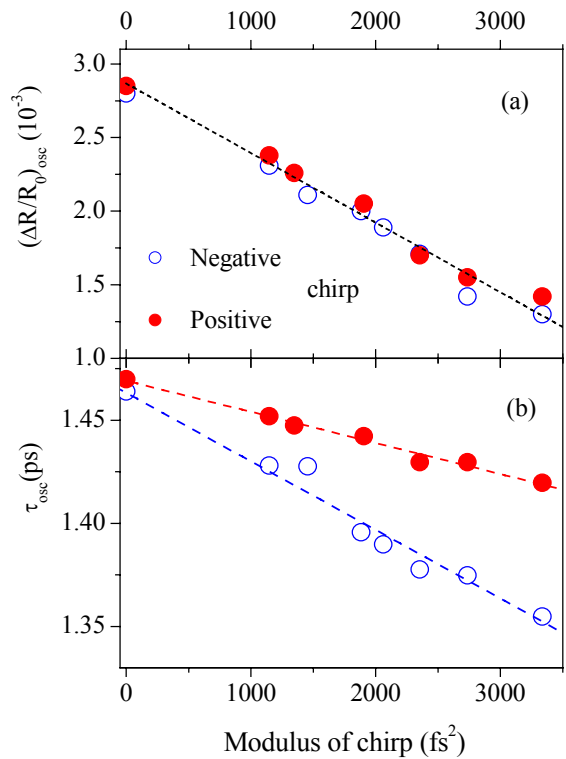

Figure 8.